\input harvmac
\input graphicx
\input color

\def\Title#1#2{\rightline{#1}\ifx\answ\bigans\nopagenumbers\pageno0\vskip1in
\else\pageno1\vskip.8in\fi \centerline{\titlefont #2}\vskip .5in}

%
%
\ifx\includegraphics\UnDeFiNeD\message{(NO graphicx.tex, FIGURES WILL BE IGNORED)}
\def\figin#1{\vskip2in}
\else\message{(FIGURES WILL BE INCLUDED)}\def\figin#1{#1}
\fi
\def\Fig#1{Fig.~\the\figno\xdef#1{Fig.~\the\figno}\global\advance\figno
 by1}
%
%
%
%

\font\ticp=cmcsc10

\def \purge#1 {\textcolor{magenta}{#1}}
\def \new#1 {\textcolor{blue}{#1}}
\def\comment#1{}

\def\\{\cr}
\def\text#1{{\rm #1}}
\def\frac#1#2{{#1\over#2}}

\def\calo{{\cal O}}

\def\roughly#1{\mathrel{\raise.3ex\hbox{$#1$\kern-.75em\lower1ex\hbox{$\sim$}}}}
\font\bbbi=msbm10 
\def\mathbb#1{\hbox{\bbbi #1}}

\def\mthsu{\mathsurround=0pt  }
\def\leftrightarrowfill{$\mthsu \mathord\leftarrow\mkern-6mu\cleaders
  \hbox{$\mkern-2mu \mathord- \mkern-2mu$}\hfill
  \mkern-6mu\mathord\rightarrow$}
\def\overleftrightarrow#1{\vbox{\ialign{##\crcr\leftrightarrowfill\crcr\noalign{\kern-1pt\nointerlineskip}$\hfil\displaystyle{#1}\hfil$\crcr}}}
\overfullrule=0pt

%
%
\lref\LIGO{
  B.~P.~Abbott {\it et al.} [LIGO Scientific and Virgo Collaborations],
  ``Observation of Gravitational Waves from a Binary Black Hole Merger,''
Phys.\ Rev.\ Lett.\  {\bf 116}, no. 6, 061102 (2016).
[arXiv:1602.03837 [gr-qc]].
}
\lref\Hawk{
  S.~W.~Hawking,
  ``Particle Creation by Black Holes,''
Commun.\ Math.\ Phys.\  {\bf 43}, 199 (1975), Erratum: [Commun.\ Math.\ Phys.\  {\bf 46}, 206 (1976)].
}
\lref\Hawkinc{
  S.~W.~Hawking,
  ``Breakdown of Predictability in Gravitational Collapse,''
Phys.\ Rev.\ D {\bf 14}, 2460 (1976).
}
\lref\Haag{R. Haag, {\sl Local quantum physics, fields, particles, algebras,} Springer (Berlin, 1996).}
\lref\Torre{
  C.~G.~Torre,
  ``Gravitational observables and local symmetries,''
Phys.\ Rev.\ D {\bf 48}, 2373 (1993).
[gr-qc/9306030].
}
\lref\AMPS{
  A.~Almheiri, D.~Marolf, J.~Polchinski and J.~Sully,
  ``Black Holes: Complementarity or Firewalls?,''
  JHEP {\bf 1302}, 062 (2013).
  [arXiv:1207.3123 [hep-th]].
}
\lref\Fuzz{
  S.~D.~Mathur,
  ``Fuzzballs and the information paradox: A Summary and conjectures,''
[arXiv:0810.4525 [hep-th]].
}
\lref\BHQIUE{
  S.~B.~Giddings,
  ``Black holes, quantum information, and unitary evolution,''
  Phys.\ Rev.\ D {\bf 85}, 124063 (2012).
[arXiv:1201.1037 [hep-th]].
}
\lref\SGmodels{
  S.~B.~Giddings,
   ``Models for unitary black hole disintegration,''  Phys.\ Rev.\ D {\bf 85}, 044038 (2012)
[arXiv:1108.2015 [hep-th]].
}
\lref\NLvC{
  S.~B.~Giddings,
  ``Nonlocality versus complementarity: A Conservative approach to the information problem,''
Class.\ Quant.\ Grav.\  {\bf 28}, 025002 (2011).
[arXiv:0911.3395 [hep-th]].
}
\lref\NVNL{
  S.~B.~Giddings,
  ``Nonviolent nonlocality,''
  Phys.\ Rev.\ D {\bf 88},  064023 (2013).
[arXiv:1211.7070 [hep-th]].
}
\lref\AMPSS{
  A.~Almheiri, D.~Marolf, J.~Polchinski, D.~Stanford and J.~Sully,
  ``An Apologia for Firewalls,''
JHEP {\bf 1309}, 018 (2013).
[arXiv:1304.6483 [hep-th]].
}
\lref\HaPr{
  P.~Hayden, J.~Preskill,
  ``Black holes as mirrors: Quantum information in random subsystems,''
JHEP {\bf 0709}, 120 (2007).
[arXiv:0708.4025 [hep-th]].
}
\lref\SGEHT{
  S.~B.~Giddings,
  ``Possible observational windows for quantum effects from black holes,''
Phys.\ Rev.\ D {\bf 90}, no. 12, 124033 (2014).
[arXiv:1406.7001 [hep-th]].
}
\lref\BCP{
  A.~Buonanno, G.~B.~Cook and F.~Pretorius,
 ``Inspiral, merger and ring-down of equal-mass black-hole binaries,''
Phys.\ Rev.\ D {\bf 75}, 124018 (2007).
[gr-qc/0610122].
}
\lref\GiShone{
  S.~B.~Giddings and Y.~Shi,
  ``Quantum information transfer and models for black hole mechanics,''
Phys.\ Rev.\ D {\bf 87}, 064031 (2013).
[arXiv:1205.4732 [hep-th]].
}
\lref\Sussxfer{
  L.~Susskind,
  ``The Transfer of Entanglement: The Case for Firewalls,''
[arXiv:1210.2098 [hep-th]].
}
\lref\BHIUE{
  S.~B.~Giddings,
 ``Black hole information, unitarity, and nonlocality,''
Phys.\ Rev.\ D {\bf 74}, 106005 (2006).
[hep-th/0605196].
}
\lref\GiShtwo{
  S.~B.~Giddings and Y.~Shi,
  ``Effective field theory models for nonviolent information transfer from black holes,''
Phys.\ Rev.\ D {\bf 89}, no. 12, 124032 (2014).
[arXiv:1310.5700 [hep-th]].
}
\lref\SGT{
  S.~B.~Giddings,
  ``Modulated Hawking radiation and a nonviolent channel for information release,''
[arXiv:1401.5804 [hep-th]], Phys.\ Lett.\ B {\bf 738}, 92 (2014).
}
\lref\SGEFT{
  S.~B.~Giddings,
  ``Nonviolent information transfer from black holes: a field theory parameterization,''
Phys.\ Rev.\ D {\bf 88}, 024018 (2013).
[arXiv:1302.2613 [hep-th]].
}
\lref\SGalg{
  S.~B.~Giddings,
  ``Hilbert space structure in quantum gravity: an algebraic perspective,''
JHEP {\bf 1512}, 099 (2015).
[arXiv:1503.08207 [hep-th]].
}
\lref\DoGi{
  W.~Donnelly and S.~B.~Giddings,
  ``Diffeomorphism-invariant observables and their nonlocal algebra,''
Phys.\ Rev.\ D {\bf 93}, no. 2, 024030 (2016).
[arXiv:1507.07921 [hep-th]].
}
\lref\HPS{
  S.~W.~Hawking, M.~J.~Perry and A.~Strominger,
  ``Soft Hair on Black Holes,''
[arXiv:1601.00921 [hep-th]].
}
\lref\tHoo{
  G.~'t Hooft,
  ``The black hole interpretation of string theory,''
Nucl.\ Phys.\ B {\bf 335}, 138 (1990).
}
\lref\BHMR{
  S.~B.~Giddings,
  ``Black holes and massive remnants,''
Phys.\ Rev.\ D {\bf 46}, 1347 (1992).
[hep-th/9203059].
}
\lref\vanR{
  M.~Van Raamsdonk,
  ``Building up spacetime with quantum entanglement,''
Gen.\ Rel.\ Grav.\  {\bf 42}, 2323 (2010), [Int.\ J.\ Mod.\ Phys.\ D {\bf 19}, 2429 (2010)].
[arXiv:1005.3035 [hep-th]].
}
\lref\EREPR{
  J.~Maldacena and L.~Susskind,
  ``Cool horizons for entangled black holes,''
Fortsch.\ Phys.\  {\bf 61}, 781 (2013).
[arXiv:1306.0533 [hep-th]].
}
\lref\Statph{
  S.~B.~Giddings,
  ``Statistical physics of black holes as quantum-mechanical systems,''
Phys.\ Rev.\ D {\bf 88}, 104013 (2013).
[arXiv:1308.3488 [hep-th]].
}
\lref\SGSB{
  S.~B.~Giddings,
 ``Hawking radiation, the Stefan-Boltzmann law, and unitarization,''
Phys.\ Lett.\ B {\bf 754}, 39 (2016).
[arXiv:1511.08221 [hep-th]].
}
\lref\GiPs{
  S.~B.~Giddings and D.~Psaltis,
  ``Event Horizon Telescope Observations as Probes for Quantum Structure of Astrophysical Black Holes,''
[arXiv:1606.07814 [astro-ph.HE]].
}
\lref\SGLIGO{
  S.~B.~Giddings,
  ``Gravitational wave tests of quantum modifications to black hole structure,''
[arXiv:1602.03622 [gr-qc]].
}
\lref\EHT{V. Fish, W. Alef, J. Anderson, K. Asada, A. Baudry, A. Broderick, C. Carilli and F. Colomer et al., ``High-Angular-Resolution and High-Sensitivity Science Enabled by Beamformed ALMA," [arXiv:1309.3519 [astro-ph.IM]].}
\lref\WABHIP{
  S.~B.~Giddings,
 ``Why aren't black holes infinitely produced?,''
Phys.\ Rev.\ D {\bf 51}, 6860 (1995).
[hep-th/9412159].
}
\lref\Susstrouble{
  L.~Susskind,
 ``Trouble for remnants,''
[hep-th/9501106].
}
\lref\BPS{
  T.~Banks, L.~Susskind and M.~E.~Peskin,
  ``Difficulties for the Evolution of Pure States Into Mixed States,''
Nucl.\ Phys.\ B {\bf 244}, 125 (1984)..
}
\lref\Harlow{
  D.~Harlow,
  ``Jerusalem Lectures on Black Holes and Quantum Information,''
Rev.\ Mod.\ Phys.\  {\bf 88}, 15002 (2016), [Rev.\ Mod.\ Phys.\  {\bf 88}, 15002 (2016)].
[arXiv:1409.1231 [hep-th]].
}
\lref\Pres{
  J.~Preskill,
  ``Do black holes destroy information?,''
In {\it Houston 1992, Proceedings, Black holes, membranes, wormholes and superstrings},
[hep-th/9209058].
}
\lref\Pagerev{
  D.~N.~Page,
  ``Black hole information,''
[hep-th/9305040].
}
\lref\Trieste{
  S.~B.~Giddings,
  ``Quantum mechanics of black holes,''
[hep-th/9412138].
}
\lref\Mathurrev{
  S.~D.~Mathur,
 ``The Information paradox: A Pedagogical introduction,''
Class.\ Quant.\ Grav.\  {\bf 26}, 224001 (2009).
[arXiv:0909.1038 [hep-th]].
}
\lref\Pageone{
  D.~N.~Page,
  ``Average entropy of a subsystem,''
Phys.\ Rev.\ Lett.\  {\bf 71}, 1291 (1993).
[gr-qc/9305007].
}
\lref\Pagetwo{
  D.~N.~Page,
  ``Information in black hole radiation,''
Phys.\ Rev.\ Lett.\  {\bf 71}, 3743 (1993).
[hep-th/9306083].
}
\lref\PHT{S.~B.~Giddings, ``Black holes, quantum information, and the foundations of physics, Physics Today {\bf 66}, 30 (2013).}

\Title{
\vbox{\baselineskip12pt  
}}
{\vbox{\centerline{Observational strong gravity and } \centerline{quantum black hole structure}
}}

\centerline{{\ticp 
Steven B. Giddings\footnote{$^\ast$}{Email address: giddings@physics.ucsb.edu}
} }
\centerline{\sl Department of Physics}
\centerline{\sl University of California}
\centerline{\sl Santa Barbara, CA 93106}
\vskip.10in
\centerline{\bf Abstract}
Quantum considerations have led many theorists to believe that classical black hole physics is modified not just deep inside black holes but at  {\it horizon scales}, or even further outward.  The near-horizon regime has just begun to be observationally probed  for astrophysical black holes -- both by LIGO, and by the Event Horizon Telescope.  This suggests exciting prospects for observational constraints on or discovery of new quantum black hole structure.  This paper overviews arguments for certain such structure and these prospects. 
\vskip.10in

\vskip.3in
\Date{}

The hundredth birthdays of general relativity (GR) and relativistic black holes mark a new watershed:  beginning of observational study of  strong gravity near black holes, initiated by the recent LIGO discovery\refs{\LIGO} and the creation of the networked Event Horizon Telescope (EHT)\EHT.  These instruments now observe effects at  near-horizon scales.  Many physicists expect that improved observations will provide increasingly precise confirmation of GR.  But a broad segment of the community investigating how black hole evolution can be consistent with quantum theory has concluded  that this requires new effects, beyond GR and local quantum field theory (LQFT), at  horizon or  larger distance scales.  Taken together, these  statements indicate  fascinating prospects for observational constraints on scenarios for quantum black hole evolution, or  observational discovery  of quantum black hole structure.

The observed black holes (BHs) range from tens to billions of solar masses, so near-horizon curvatures are tiny, and a first expectation is that  it's preposterous that any new quantum effects could be relevant.  However, BHs must be describable in a quantum framework, and such attempts\refs{\Hawk,\Hawkinc} have led to a foundational crisis in physics, sometimes called the ``information paradox."  As a result, many theorists now consider modifications to a classical BH picture extending to the event horizon or beyond.

While debate continues, the basic argument is that quantum information that falls into a BH can't escape because of locality, can't be preserved in  microscopic remnants without implying catastrophic instabilities, and  can't be lost, violating quantum mechanics, without  disastrous energy nonconservation.  These statements follow from LQFT's prohibition of transfer of information outside the light cone, the  statement that the unbounded number of species of microscopic remnants required would yield unbounded-production instabilities\refs{\WABHIP,\Susstrouble}, and a connection between information loss and energy non-conservation spelled out in \refs{\BPS}.  (For reviews see \refs{\Pres\Pagerev\Trieste\Mathurrev\PHT-\Harlow}.)
Increasingly many researchers conclude that BH evolution must respect quantum mechanics and specifically unitarity, but then the central question is how this can happen.   This represents a  ``unitarity crisis" in  fundamental physics.

Let us explore the assumption that quantum mechanics must ultimately be respected.  The trouble arises because one can lose quantum information to the internal state of a BH; if this then evaporates, without emitting its information, unitarity is  violated.  Put differently, evolution produces entanglement between the  BH and its environment.  This occurs if one member of an EPR pair falls into a BH, or naturally via the Hawking process, where Hawking particles and their partner excitations inside the BH are entangled.  Unitarity requires this entanglement to transfer out\refs{\HaPr\GiShone-\Sussxfer}.  But locality of LQFT forbids this:  no quantum information transfers outside the light cone.

Here I implicitly assume that one can decompose the BH interior and environment into distinct quantum subsystems.  This is subtle in gravity: the usual LQFT approach to this is based on existence of local operators\refs{\Haag}.   In GR local gauge (diffeomorphism)-invariant operators don't exist\refs{\Torre}.  One can see explicitly that the ``gravitational dressing" needed to satisfy the constraint equations obstructs commutativity of gauge-invariant operators outside the light-cone\refs{\SGalg,\DoGi}.  Specifically, a particle is inseparable from its gravitational field, and the long-range field part of the operator that creates both together impedes commutativity.  While ideas exist for using such gravitational dressing to ultimately store or communicate  information\refs{\tHoo,\HPS}, many do not understand how this can be a big enough effect for BHs.   We will therefore assume an approximate decomposition, and investigate necessary constraints on interactions between these subsystems, whether due to such effects, or due to new quantum effects.

Preservation of unitarity requires transfer of information from the BH interior to outside, so over scales of size $R$, the horizon radius.  While this is forbidden by locality, we have found that locality in quantum gravity is subtle.  Partly for this reason, the theory community has increasingly considered the idea that there must be new effects not respecting usual LQFT -- extending to scales $\roughly> R$.  While this essential proposal is older\refs{\BHMR}, it has been famously realized recently in the ``firewall" scenario\AMPS, where  a BH transitions to a new object with information residing at the would-be horizon.   Viewed from the perspective of the semiclassical BH geometry, this transfer of information to the horizon is nonlocal. Apparently more radical modifications of LQFT, on even greater scales, have alternately been considered\refs{\vanR,\EREPR}, partly connected with ideas that spacetime  emerges from some more fundamental structure.  

Firewalls represent  dramatic breakdown of spacetime at the would-be horizon; we should consider  the possibility of more conservative ways to save quantum mechanics.   Transfer of information from the BH subsystem to outside is needed, apparently requiring new interactions not respecting the LQFT locality.  The firewall scenario assumed  transfer  across scales $\sim R$ -- larger than the Earth-Neptune distance, for the biggest known BHs -- which then abruptly halts at a microscopic distance from the horizon.  It appears more natural to assume that this transfer is over scales set by $R$, but is less abrupt -- so the new interactions  transfer information to the immediate vicinity of the horizon\refs{\SGmodels,\BHQIUE}, or ``atmosphere." 

To model such ``soft" transfer, we conservatively assume that the new interactions are described as minimal corrections to LQFT; this for example can minimize the effect on an infalling observer. Then, the needed interactions are parameterized  as couplings between the BH internal state and the fields of the atmosphere\refs{\SGEFT,\GiShtwo}.  There are other constraints to consider.  First, these couplings must be able to transfer of order one qubit per light-crossing time $R$ -- a scale set by the evaporation rate -- to save unitarity.  This can for example be understood from arguments by Page\refs{\Pageone,\Pagetwo}; after an approximate mid-point of black hole evaporation, the von Neumann entropy of the black hole has to begin to decrease at a rate $\sim 1/R$ in order to  ultimately reach zero when the black hole evaporates.  This could be achieved by simply coupling the BH state to the photon or other fields through interaction terms of sufficient strength to, {\it e.g.}, emit $\calo(1)$ quantum per time $R$.  

But,  general such couplings apparently spoil the beautiful story of BH thermodynamics \refs{\SGmodels,\BHQIUE,\AMPS,\Statph}, with Bekenstein-Hawking entropy given by the horizon area.  The couplings provide an extra channel for information and energy to be emitted from the BH, so if one tries to bring a BH into equilibrium with a thermal bath, they  violate detailed balance.  This, for example, would arise by the BH emitting more radiation than it absorbs, or emitting more of specific particle species than others, due to these extra couplings with the surrounding fields.  To avoid this, one needs universal couplings to all fields, and such that these couplings yield emission rates governed by the Hawking temperature $T_H$.  Coupling to the stress tensor $T_{\mu\nu}$ for all fields, via an action term
\eqn\stresscoup{\Delta S=\int_{\rm atmosphere} d^4x \sqrt{-g} H^{\mu\nu}(x) T_{\mu\nu}(x)\ ,}
where $H^{\mu\nu}(x)$ is an operator with non-trivial matrix elements between BH quantum states, satisfies universality. The  operator $H^{\mu\nu}(x)$ couples the BH states to the nearby fields; it is not itself a quantum field.  This universal coupling also addresses Gedanken experiments involving BH mining\refs{\AMPS,\SGT}.  ``Softness" is incorporated if $H^{\mu\nu}(x)$ has variation on scales $\sim R$, not more rapidly. 

In \stresscoup, the quantum average $\langle H^{\mu\nu}(x)\rangle$ behaves like a quantum correction to the classical metric.  This coupling has a job to do -- relay $\calo(1)$ qubit from the BH per $R$ time; this is achieved with magnitude\refs{\SGT} $\langle H^{\mu\nu}(x)\rangle = \calo(1)$.  This can be alternately explained by noting that unitarity restoration requires an  $\calo(1)$ modification of the outgoing Hawking radiation, which arises from  $\calo(1)$ variation in the metric.  

The arguments for this model can and should be sharpened, but lead to an interesting proposal:  ``minimal" effects needed to restore unitarity and preserve BH 
thermodynamics\refs{\SGSB} have effective description as  {\it strong but soft} quantum fluctuations in the metric.  These yield characteristic curvatures ${\cal R}\sim 1/R^2$, so are nonviolent to an infalling observer.  But, they {\it can} alter observations of the BH atmosphere, {\it e.g.} by producing $\calo(1)$ deflections in geodesics.

EHT and LIGO have the potential to constrain or discover this or other quantum structure of BHs.   EHT observes light emitted by matter accreting into Sgr A$^*$, at the center of our galaxy; general relativity predicts an image with a characteristic shadow, and a bright photon ring just outside it.  The latter arises from an accumulation of geodesics near the light orbit of the BH, and both features are expected to be significantly altered if new metric perturbations extend past this orbit, at $3R/2$ for Schwarzschild\refs{\SGEHT}.  Checking this involves parameterizing\refs{\SGEHT,\GiPs} such fluctuations and numerical ray tracing of photons from accreting matter\refs{\GiPs}.

Gravity waves observed by LIGO\LIGO\ also access this domain.  As inspiraling BHs ``plunge" and merge -- at and before the gravity wave peak -- they probe each other's near-horizon geometry\refs{\SGLIGO}.  So far, combination of the residuals shown in \LIGO\ does not reveal significant deviations from GR.  This may  even rule out some near-horizon modifications, such as ``hard" (high-momentum) perturbations that one might expect from fuzzballs\refs{\Fuzz}.  More precise constraints require both parameterization of proposed deviations, and their numerical evolution in this regime, where also  a significant part of the perturbation can fall into the final BH.

But the message is clear: many theorists are now exploring horizon-scale quantum modifications to black holes expected to be necessary for unitary evolution --  and EHT and LIGO are now observing these regimes.  This offers the exciting possibility to constrain or observe  profoundly new effects apparently needed to rescue quantum mechanics from black holes, and is worthy of continued effort.

\bigskip\bigskip\centerline{{\bf Acknowledgments}}\nobreak

This work  was supported in part by the Department of Energy under Contract No. DE-SC0011702, and by Foundational Questions Institute (fqxi.org) 
grant FQXi-RFP-1507.

\listrefs
\end